%% file: ImmersiveDesign_Main.tex
\documentclass[preprint,journal]{vgtc}                

\ifpdf
  \pdfoutput=1\relax                   
  \usepackage{graphicx}                
  \DeclareGraphicsExtensions{.pdf,.png,.jpg,.jpeg} 
\else
  \usepackage{graphicx}                
  \DeclareGraphicsExtensions{.eps}     
\fi%

\graphicspath{{figures/}{pictures/}{images/}{./}} 

\usepackage{microtype}                 
\PassOptionsToPackage{warn}{textcomp}  
\usepackage{textcomp}                  
\usepackage{mathptmx}                  
\usepackage{times}                     
\usepackage{cite}                      
\usepackage{tabu}                      
\usepackage{booktabs}                  
\usepackage{enumitem}

\usepackage{color}
\usepackage{amssymb}
\usepackage{array}
\usepackage{multirow}
\usepackage{dblfloatfix}
\usepackage[table]{xcolor}

\newcommand{\etal}{et al.~}
\newcommand{\eg}{e.g.,~}

\definecolor{myred}{RGB}{128,57,65}
\definecolor{myblue}{RGB}{49,71,112}
\definecolor{myorange}{RGB}{186,110,50}
\definecolor{myyellow}{RGB}{189,150,21}
\definecolor{myyellow}{RGB}{189,150,21}
\definecolor{dullblue}{RGB}{168,185,169}

\newcommand{\communal}{\textbf{Communal}}
\newcommand{\personal}{\textbf{Personal}}
\newcommand{\Active}{\textbf{Active}}
\newcommand{\emergent}{\textbf{Emergent}}

\onlineid{0}

\vgtccategory{theory/model}
\vgtcpapertype{please specify}

\title{Design by Immersion: A Transdisciplinary Approach to Problem-Driven Visualizations}

\author{Kyle Wm. Hall, Adam J. Bradley, Uta Hinrichs, Samuel Huron, Jo Wood \\ Christopher Collins, and Sheelagh Carpendale}
\authorfooter{
\item Kyle Wm. Hall is with Temple University, Philadelphia, PA, USA. E-mail: k.wm.hall@temple.edu.
\item Adam J. Bradley is with Ontario Tech University, Oshawa, ON, Canada. E-mail: adam.bradley@uoit.ca.
\item Uta Hinrichs is with the University of St Andrews, Fife, United Kingdom. E-mail: uh3@st-andrews.ac.uk.
\item Samuel Huron is with the T\'{e}l\'{e}com Paristech, Universit\'{e} Paris-Saclay, Paris, France. E-mail: samuel.huron@telecom-paristech.fr.
\item Jo Wood is with City, University of London, London, United Kingdom. E-mail: j.d.wood@city.ac.uk.
\item Christopher Collins is with Ontario Tech University, Oshawa, ON, Canada. E-mail: christopher.collins@uoit.ca.
\item Sheelagh Carpendale is with the University of Calgary, Calgary, Alberta, Canada, and with Simon Fraser University, Burnaby, British Columbia, Canada. E-mail: sheelagh@ucalgary.ca.
}

\ieeedoi{10.1109/TVCG.2019.2934790}

\abstract{While previous work exists on how to conduct and disseminate insights from problem-driven visualization projects and design studies, the literature does not address how to accomplish these goals in transdisciplinary teams in ways that advance all disciplines involved. In this paper we introduce and define a new methodological paradigm we call \textit{design by immersion}, which provides an alternative perspective on problem-driven visualization work. Design by immersion embeds transdisciplinary experiences at the center of the visualization process by having visualization researchers participate in the work of the target domain (or domain experts participate in visualization research). Based on our own combined experiences of working on cross-disciplinary, problem-driven visualization projects, we present six case studies that expose the opportunities that design by immersion enables, including (1)~exploring new domain-inspired visualization design spaces, (2)~enriching domain understanding through personal experiences, and (3)~building strong transdisciplinary relationships. Furthermore, we illustrate how the process of design by immersion opens up a diverse set of design activities that can be combined in different ways depending on the type of collaboration, project, and goals. Finally, we discuss the challenges and potential pitfalls of design by immersion.%
} 


\CCScatlist{ 
 \CCScat{K.6.1}{Management of Computing and Information Systems}%
{Project and People Management}{Life Cycle};
 \CCScat{K.7.m}{The Computing Profession}{Miscellaneous}{Ethics}
}

\vgtcinsertpkg

\begin{document}
\firstsection{Introduction}
\maketitle


\input{01_introduction}

\input{02_related-work}

\input{03_case-studies}

\input{04_opportunities.tex}

\input{05_immersion-process.tex}

\input{07_outlook.tex}

\input{08_conclusions.tex}

\acknowledgments{This research was supported in part by the Natural Sciences and Engineering Research Council of Canada (NSERC), Alberta Innovates Technology Futures (AITF), and SMART Technologies ULC. K. Wm. Hall thanks NSERC for its support through the Vanier Canada Graduate Scholarships Program.
}

\bibliographystyle{abbrv-doi}

\bibliography{template}
\end{document}

%% file: 01_introduction.tex
In 1986, McCormick et al.~\cite{McCormickEtAl1987} advocated that scientists, engineers and visualization researchers should form collaborative teams such that domain needs and processes provide a basis for solving visualization challenges. However, realizing this vision is complex as there are multiple paradigms for research involving different disciplines. Kirby and Meyer~\cite{KirbyMeyer-2013} characterize multidisciplinary work as addressing challenges that, while being associated with specific domains, require expertise from multiple disciplines. In the multidisciplinary paradigm, ``researchers work in parallel with clearly defined roles and specific tasks that provide added benefit to their disciplinary goal''~\cite[p.83]{KirbyMeyer-2013}. They describe the interdisciplinary research paradigm as addressing problems lying outside disciplinary confines, requiring the intersection of multiple disciplines. In this paper, we present a methodology, \textit{design by immersion}, that is based on a third paradigm we characterize as \textit{transdisciplinarity} where the lines between visualization researchers and domain experts blur as individuals move beyond working in a single domain. The immersive designer works---partially or fully---in both their home discipline (visualization or domain) and the `other' discipline (domain or visualization). Design by immersion has many benefits. It can facilitate collaboration and accelerate project development by building trust and deepening the dialogue between collaborators. From a visualization perspective, it expands the portfolio of existing visualization design processes in ways that encourage active participation of domain experts in the visualization process, allowing for the fluid integration of visualization processes and domains as well as novel perspectives on visualization. Design by immersion also fosters personal development, such as the acquisition of new skills and experiences, enabling a better understanding of different research perspectives and practices. We do not claim that design by immersion is ``better'' than existing practises, but it can offer new ways of looking at visualization design. Design by immersion is well suited to problem-driven visualization work.

In contrast to technique-driven visualization which aims to create ``new and better techniques without necessarily establishing a strong connection to a particular documented user need'', the goal of problem-driven visualization is ``to work with real users to solve their real-world problems''~\cite[p.2432]{SedlmairEtAl2012}. However, problem-driven visualization research comes with challenges introduced by gaps in both knowledge bases and cultures~\cite{Kerzner2018, Munzner2009, SeinEtAl2011, SedlmairEtAl2012, McKennaEtAl2014, McCurdyEtAl2016, Muller2007}. In response to these challenges, the visualization community has developed guidelines for problem-driven and multidisciplinary visualization projects~\cite{WongBELIV2018,SedlmairEtAl2012,SimonEurovis2015}. Specific design and workshop activities~\cite{Kerzner2018,McKennaEtAl2014}, visualization models~\cite{Munzner2009}, and collaborative paradigms~\cite{SimonEurovis2015, Wijk2006} have been explored for engaging with domain experts. However, as Wood et al. note, visualization literature generally creates an opposition between visualization and domain experts~\cite{WoodEtAl2014}. For example, in their nine-stage design study methodology framework, Sedlmair \etal advocate that researchers should clearly identify collaborators' roles prior to characterizing a domain and engaging in the design process~\cite{SedlmairEtAl2012}. Similarly, action design research~\cite{SeinEtAl2011} (suggested by McCurdy \etal\cite{McCurdyEtAl2016} as a visualization design framework) advocates for clearly assigning roles in collaborative problem-driven design projects. In contrast, a growing number of visualization case studies report a blurring of the boundaries between visualization and target domains~\cite{HinrichsEtAl2016, HinrichsEtAl2017, WoodEtAl2014, IsenbergEtAl2008, BradleyEtAl2016}. In these instances, the roles of the researchers involved cannot be distinctly classified and may have even shifted over the course of the collaboration. Even though Sedlmair \etal\cite{SedlmairEtAl2012} call for role definition, they concede that problem-driven visualization work may involve a single person in the role of both visualization and domain expert. Similarly, Wong et al.~\cite[p.1]{WongBELIV2018} note that visualization ``tools historically required the users to not only be \textit{domain experts}, i.e., have expertise in a specific discipline, but also have the time and motivation to become visualization experts.'' 

While previous work hints at the benefits of transdisciplinary approaches to problem-driven visualization work, methodologies to facilitate such work are absent in the literature. Using our own experiences working on distinct transdisciplinary visualization projects, we introduce design by immersion as a paradigm that captures and supports transdisciplinary approaches to problem-driven visualization where visualization experts immerse themselves in a target domain in order to inform visualization processes, and/or domain experts actively engage in visualization design processes to help explore and define visual solutions to their real-world problems. 

Design by immersion is similar to cultural immersion---the direct experiencing of and engagement with communities, environments, and/or languages that are different from one's own. Work in education, cultural studies, sociology, and ethnography has shown that cultural immersion enables a first-hand experience of the target community/environment/language that remote studies cannot provide~\cite{Nieto_2006}. Cultural immersion can lead to a more nuanced understanding of the target scenario's characteristics, corresponding processes, and challenges, and to an increased awareness of one's own assumptions and biases in relation to the target domain. 

Design by immersion is connected to experiential learning. As Kolb states: ``Knowledge results from the combination of grasping and transforming experience''~\cite[p.51]{Kolb2015} and ``[experiential] Learning is the process whereby knowledge is created through the transformation of experience.''~\cite[p.49]{Kolb2015}. Much has been written about the positive impacts of experiential learning~\cite{GilinYoung2009,Hu2016,Lee2008,LiskoODell2010,MooneyEdwards2001}. Transdisciplinary experiences offer researchers opportunities to encounter and learn more richly about other domains and to transfer this knowledge to visualization design. 

Leveraging the ideas of cultural immersion and experiential learning, we consider design by immersion as a new way to describe and guide collaborative visualization research. In this paper, we illustrate how design by immersion can be applied to a broad range of collaborative problem-driven visualization scenarios using a number of case studies. We discuss the opportunities it supports, including: exploring new domain-inspired visualization design spaces (e.g., new domain-focused charts in Case Study~\#6); enriching domain understanding through personal experiences (e.g., development of a PhD thesis in Case Study~\#5); and building strong transdisciplinary relationships (e.g., ongoing collaborative publications in Case Study~\#4). We provide hands-on guidance on how to engage in design by immersion with concrete, modular design activities that can be tailored to different types of projects. Finally, we reflect on the potential challenges of this method. 

This paper contributes: \textbf{(1)} a new methodology to support transdisciplinary problem-driven visualization research, \textbf{(2)} an illustration of design by immersion using a range of different real-world visualization case studies with a discussion of the opportunities that design by immersion introduces to visualization research in general, and \textbf{(3)} guidance for approaching design by immersion in visualization research with a discussion of potential challenges.

%% file: 02_related-work.tex
\section{Related Work}
\label{sec:related-work}
Our work builds and expands on previous research in problem-driven visualization and design studies. Design by immersion is related to grounded evaluation~\cite{IsenbergEtAl2008} and pre-design empiricism~\cite{BrehmerEtAl2014}, both of which advocate the use of exploratory qualitative studies to inform design. However, the activities and themes in our approach go beyond treating the other domain as the object of study.

\subsection{Frameworks to Guide Problem-Driven Visualization}
A number of frameworks and guidelines attempt to systematize the process of problem-driven visualization, conducting design studies, and working with domain experts. Munzner's nested model~\cite{Munzner2009} deconstructs problem-driven visualization design into four components (domain problem and data characterization; operation and data type abstraction; visual encoding and interaction design; and algorithm design) while emphasizing evaluation. Design study methodology was described in the nine-stage framework by Sedlmair et al.~\cite{SedlmairEtAl2012}, which provides the high-level stages of a design study. Their work and discussions highlight the multidisciplinary nature of visualization design studies. The Design Activity Framework~\cite{McKennaEtAl2014} contributes structure to the process by breaking visualization design down into a set of activities (understand, ideate, make, deploy) that consist of motivations, methods, and outcomes. Wong et al.~\cite{WongBELIV2018} present a characterization of domain experts, features of visualization systems for those experts, and corresponding design guidelines. However, visualization literature tends to separate visualization and domain experts into explicitly distinct groups~\cite{Munzner2009, SedlmairEtAl2012, McKennaEtAl2014, WongBELIV2018}. There is often an implicit or explicit assumption that the visualization researcher designs a visualization \textit{for} a particular domain (or problem) \cite{vandeMoerePurchase2011, Kindlemann2014, rind2016, house2005}. Such perspectives do not capture the possibility and potential benefits of design by immersion---a deeply collaborative visualization design process where domain-inspired solutions arise from transdisciplinary practices and contribute to all disciplines involved. 

Nevertheless, design by immersion aligns with and extends existing themes in the literature. For example, Simon~\etal \cite{SimonEurovis2015} introduce the design study Liaison role in which an individual team member (either a domain or visualization expert) with additional knowledge in the ``other'' discipline facilitates visualization design by, in part, serving as a knowledge conduit between disciplines. There are similarities between Liaison-supported design studies and design by immersion, and an immersed researcher is well positioned to serve as a Liaison. Transitioning to design by immersion involves shifting from a role-based paradigm with an individual bridging the separate domain and visualization spheres to a collaborative process that brings together the two spheres. Design by immersion can be considered as a broader transdisciplinary approach and mindset that aims at collaboratively identifying and leveraging synergies between the domain and visualization spheres, allowing visualization design processes and roles to fluidly evolve, eventually blurring disciplinary boundaries.

\subsection{Action Design Research}
Considering design research from beyond the field of visualization, strategies exist that offer a wider view of the role of the target domain in visualization design. For example, Action Design Research (ADR) approaches design from the perspective that technological artifacts represent both design knowledge (visualization theory) and design context (target domain knowledge and influences from users)~\cite{SeinEtAl2011,McCurdyEtAl2016}. ADR emphasizes the interconnected nature of: 1)~building tools, 2)~intervening in the target domain via these tools, and 3)~evaluating what has been built. In ADR, these tasks are tightly bound in successive build-intervene-evaluate cycles where ``evaluation is \textit{not} a separate stage of the research process that follows building''~\cite[p.43]{SeinEtAl2011}. However, ADR promotes an artifact-centric perspective to design with a particular focus on evolution as well as target domain intervention and disruption. It does not explore transdisciplinary opportunities and their impact.

\subsection{Participatory Design}
Designers are increasingly focused on including users and stakeholders in the design process. For example, participatory design (\eg~\cite{schuler1993participatory}) 
is explicitly multidisciplinary and collaborative. Participatory design has been used by a number of visualization researchers (\eg~\cite{Henry2006,Chevalier2010,roberts2016sketching,Mendez2018,lloyd2011human}), and Sanders \etal have previously proposed a framework to organize participatory design tools and processes~\cite{sanders2010framework}. Based on Muller's survey~\cite{Muller2007} of participatory design approaches, its characteristic interdisciplinarity (what Muller calls \textit{hybridity}) stems from settings, activities and artifacts that encourage the creation of interdisciplinary spaces where designers and users meet to discuss and actively work through potentially differing perspectives. In the context of visualization, participatory design varies substantially from discussions with domain experts~\cite{Henry2006,Chevalier2010}, through potential users sketching design solutions~\cite{roberts2016sketching,Mendez2018} and using real domain data as a mediator in data-driven wireframes and prototypes~\cite{lloyd2011human}, to domain experts creating paper prototypes of their ideas~\cite{Henry2006}. 
While design by immersion aligns with ideas of stakeholder involvement as in participatory design, it also goes beyond them. In design by immersion, collaborations between stakeholders and designers shift from being structured through interdisciplinary spaces, artifacts and decisions to involving the personal acquisition of skills and field expertise in the visualization domain and/or the stakeholders' domain. Design by immersion emphasizes the transdisciplinary transformation of individuals and the opportunities these transformations present for design. By engaging potential users and stakeholders in the design process, participatory design invites crossover between domains, and could serve as a starting point for transitioning to design by immersion.

\subsection{Transdisciplinary Visualization Work}
Design by immersion also relates to ideas from the digital humanities where visualization has started to play an increasingly important role as a new methodology~\cite{Jaenicke_2016}. Hinrichs et al. have discussed the role of visualization as a \textit{mediator} between visualization and humanities researchers, and as a transdisciplinary \textit{speculative process} advancing all disciplines involved~\cite{Hinrichs_2018a}. This work emphasizes the importance of considering visualization as a process that not only enables communication between different disciplines, but also allows a collaborative reflection on assumptions inherent in each discipline. Our work builds on this research by defining and situating design by immersion in the broader transdisciplinary context of problem-driven visualization.

%% file: 03_case-studies.tex
\newcommand*{\belowrulesepcolor}[1]{%
  \noalign{%
    \kern-\belowrulesep
    \begingroup
      \color{#1}%
      \hrule height\belowrulesep
    \endgroup
  }%
}
\newcommand*{\aboverulesepcolor}[1]{%
  \noalign{%
    \begingroup
      \color{#1}%
      \hrule height\aboverulesep
    \endgroup
    \kern-\aboverulesep
  }%
}

\definecolor{platinum}{rgb}{0.9, 0.89, 0.89}

\begin{table*}[t!]
\caption{Activities and their connections to the opportunities of design by immersion. Activities are classified as \textbf{D}ata analysis, \textbf{S}tudy methods, \textbf{P}rototyping, \textbf{L}earning about the domain, and \textbf{C}ommunicating with collaborators.  
Numbers in the three columns are linked to the case studies.  } 
\vspace{-1.2em}
\begin{center}
\begin{tabular}{ m{0.08\columnwidth}m{1.12\columnwidth}m{0.25\columnwidth}m{0.18\columnwidth}m{0.18\columnwidth}}
\multicolumn{2}{c}{\centering \textbf{Design by Immersion Activities}} & \centering \textbf{Enrich Domain Understanding} & \centering \textbf{Explore\\ New Spaces}& \centering \textbf{Build \\Relationships} \tabularnewline \midrule 

\belowrulesepcolor{platinum}
\arrayrulecolor{lightgray} \rowcolor{platinum}
\centering \textbf{\textit{D-1}} & Undertake domain-specific data analysis independently  & \centering $1,4,5,6$ & & \centering $5$ \tabularnewline 
\aboverulesepcolor{platinum}

\midrule 
\belowrulesepcolor{platinum}
\rowcolor{platinum}
\centering \textbf{\textit{D-2}} & Enrich datasets meaningfully by deriving new data 
& \centering $5$ & \centering $4, 5$ & \centering $4$ \tabularnewline 
\aboverulesepcolor{platinum}

\midrule \belowrulesepcolor{platinum}
\rowcolor{platinum}
\centering \textbf{\textit{D-3}} & Analyze data collaboratively with domain experts
& \centering $3,4,5,6$ & & \centering $3,4,5,6$ \tabularnewline 
\aboverulesepcolor{platinum}

\midrule
\centering \textbf{\textit{S-1}} & Observe domain experts practices unobtrusively in situ
& \centering $2,6$ & & \tabularnewline 

\midrule
\centering \textbf{\textit{S-2}} & Keep documentation of experiences
& \centering $1,6$ & \centering & \centering $4$\tabularnewline

\midrule
\centering \textbf{\textit{S-3}} & Interview collaborators  
& \centering 2 & \centering  & \centering \tabularnewline 

\midrule
\centering \textbf{\textit{S-4}} & Attend meetings and discussions in the other domain  
& \centering $2,5$ & \centering $5$ & \centering $5$ \tabularnewline

\midrule \belowrulesepcolor{platinum}
 \rowcolor{platinum}
\centering \textbf{\textit{P-1}} & Develop visualizations in the context of evolving collaborative research projects with multiple disciplines & \centering $5$ &\centering $4,5$ & \centering $4,5$ \tabularnewline 
\aboverulesepcolor{platinum}

\midrule \belowrulesepcolor{platinum}
 \rowcolor{platinum}
\centering \textbf{\textit{P-2}} & Develop visual encodings that explicitly incorporate and take inspiration from domain knowledge and practices
&  \centering $3$& \centering $4,6$ & \tabularnewline 
\aboverulesepcolor{platinum}

\midrule \belowrulesepcolor{platinum}
 \rowcolor{platinum}
\centering \textbf{\textit{P-3}} & Start ideating early in the design process using pre-existing domain knowledge
& \centering $5$ & \centering $1,3,4,5$ & \tabularnewline 
\aboverulesepcolor{platinum}

\midrule \belowrulesepcolor{platinum}
 \rowcolor{platinum}
\centering \textbf{\textit{P-4}} & Iterate rapidly and collaboratively on designs by leveraging informal domain expert feedback
& \centering $5$ & \centering $1,4,5$ & \tabularnewline 
\aboverulesepcolor{platinum}

\midrule \belowrulesepcolor{platinum}
 \rowcolor{platinum}
\centering \textbf{\textit{P-5}} & Self-critique designs from the visualization and domain perspectives
& & \centering $3,4,5,6$ & \tabularnewline 
\aboverulesepcolor{platinum}

\midrule
\centering \textbf{\textit{L-1}} & Engage directly with both domain-specific and visualization literature
&\centering $3,4,5,6$ & \centering $3,5$ & \centering $3,5,6$ \tabularnewline 

\midrule
\centering \textbf{\textit{L-2}} & Gain broader exposure to domain concepts beyond domain problem
& \centering $5,6$ &  \centering $5,6$ & \centering $5$ \tabularnewline 

\midrule
\centering \textbf{\textit{L-3}} & Establish domain-based design considerations for visualizations
&\centering $5$ & \centering $5,4,6$ & \centering $5$\tabularnewline 

\midrule
\centering \textbf{\textit{L-4}} & Receive informal training from collaborators
& \centering $1$ & \centering & \centering $5$ \tabularnewline

\midrule
\centering \textbf{\textit{L-5}} & Participate in simulations of domain work  
& \centering $2$ & \centering  & \centering  $2$ \tabularnewline 

\midrule \belowrulesepcolor{platinum} \rowcolor{platinum}
\centering \textbf{\textit{C-1}} & Use language that resonates with collaborators
&\centering $3,5$ & \centering $4$ & \centering $4,5,6$ \tabularnewline 
\aboverulesepcolor{platinum}

\midrule \belowrulesepcolor{platinum} \rowcolor{platinum}
\centering \textbf{\textit{C-2}} & Relate across disciplines through common 
knowledge and experiences 
& & \centering $5$& \centering $1,4,5,6$ \tabularnewline 
\aboverulesepcolor{platinum}

\midrule \belowrulesepcolor{platinum} \rowcolor{platinum}
\centering \textbf{\textit{C-3}} & Engage in informal peer-to-peer communication with domain experts about domain science and visualizations
& \centering $4,5,6$ & \centering $4,5,6$ & \centering $4,5,6$ \tabularnewline 
\aboverulesepcolor{platinum}

\midrule \belowrulesepcolor{platinum} \rowcolor{platinum}
\centering \textbf{\textit{C-4}} & Translate concepts and material for design team members coming from \newline predominantly visualization or target domain backgrounds
&  \centering $3$ & &  \centering $3,6$ \tabularnewline 
\aboverulesepcolor{platinum}

\midrule \belowrulesepcolor{platinum} \rowcolor{platinum}
\centering \textbf{\textit{C-5}} & Brainstorm with collaborators about methods that would best elicit implicit knowledge & \centering $2$ & \centering $2$ & \centering $2$ \tabularnewline 
\aboverulesepcolor{platinum}

\arrayrulecolor{black} \bottomrule
\end{tabular}
\end{center}
\label{table:activities} 
\vspace{-2.5em}
\end{table*}

\begin{figure}[t] 
\centering
\includegraphics[width=0.85\columnwidth]{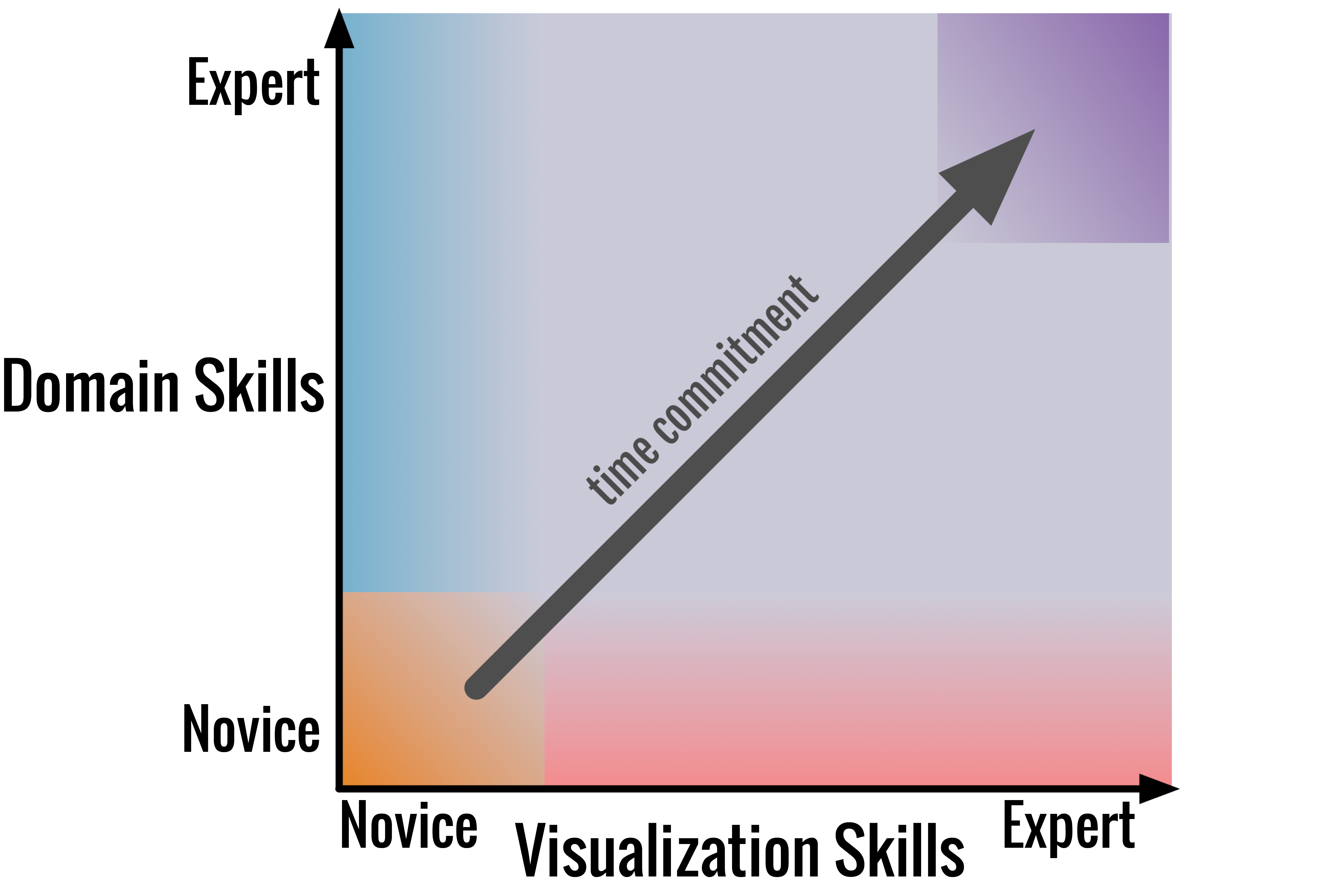} 
\vspace{-0.8em}
\caption{\textit{The immersive skills space.} Immersion can result in improving skills in both visualization and domain areas. Traversal of this space from novice to expert involves choosing where and how much time to commit to skill acquisition along the two dimensions.}
\label{fig:immersion-approaches} 
\vspace{-1.5em}
\end{figure}

\section{Design by Immersion}
\label{sec:immersive-def}
This paper is motivated by our---the authors'---experience of transdisciplinarity in our own individual collaborative visualization design projects. We collectively found it difficult to connect these experiences to the design processes and strategies in the visualization literature. While working on different projects, engaging with different domains, all of us had similar experiences with immersion. Reflecting on and discussing our design approaches, we found that all of us had in some way engaged with activities typical of the ``other'' domain and, in this way,  drifted towards becoming members of this ``other'' community. We all found this immersion to be challenging but also transformative and enriching to our projects and our own perspectives on research. While our approaches and experiences took on different forms, taken together they illustrate a unique transdiciplinary approach to problem-driven visualization we define as design by immersion. 

\textbf{Design by Immersion} \textit{is a methodology for problem-driven visualization design where visualization researchers (or target domain experts) engage with and participate in the work of another domain such that visualization design, solutions, and knowledge emerge from these transdisciplinary experiences and interactions.}

This definition is based on our own transdisciplinary visualization projects as discussed below in Section~\ref{sec:immersion} to illustrate and explicate design by immersion. We intentionally chose this methodology of characterizing design by immersion based on a small yet diverse number of exemplary visualization case studies that we know well as this enabled us to discuss the nuances of this design approach. While indications of design by immersion may be present in previous design studies, extracting traces of this approach from such work would be difficult without insider knowledge of these studies. We expect that future instantiating of design by immersion will expand perspectives on and understanding of design by immersion.

From a visualization perspective, design by immersion means to immerse oneself in the target domain and to engage with the data and analysis processes in the ways that domain experts do, to inform visualization processes and design. From the perspective of a domain expert, it means to engage with visualization as a design and thinking process in order to help explore and define approaches or solutions to a problem. This method is a flexible approach to problem-driven visualization design that can take on many forms, depending on the disciplines and types of collaborators.

Our definition exhibits four themes, and we use them as lenses to connect our case studies to our design by immersion definition.

    \textit{\textbf{\communal.}}~Researchers enter into each others' domains, and \indent existing communities with their own actors and cultures. 
    
    \textit{\textbf{\personal.}}~Researchers become intimately concerned with, affected \indent by and personally involved in the other domain. 

    \textit{\textbf{\Active.}}~Researchers actively engage in the other domain, \indent participating in domain activities, not just observing activities.

    \textit{\textbf{\emergent.}}~The processes and results of this approach have their \indent origins in and emerge from transdisciplinary interactions between \indent visualization and the target domain.

A key characteristic of design by immersion is that researchers transform and enrich their knowledge and skills through transdisciplinary experiences as demonstrated in Figure~\ref{fig:immersion-approaches}. Each discipline is represented as an axis. The expertise of collaborators in their own ``home'' discipline (visualization or target domain) can vary, as represented in these axes that span from ``novice'' to ``expert''. There is variation among visualization researchers (and domain experts) in terms of expertise, such as graduate students vs. senior researchers. It is common for visualization experts to have minimal knowledge or expertise in the target domain, and vice versa. These boundary cases are represented by the red and blue rectangles (see Figure~\ref{fig:immersion-approaches}). Of course, there are people who are novices in both the target domain and visualization, corresponding to the lower bound on the knowledge space (the orange square in Figure~\ref{fig:immersion-approaches}). Design by immersion results in increasing one's skills and knowledge in disciplines other than one's own. Maximum time commitment can even result in becoming a  \textit{dual citizen} (the purple square in Figure~\ref{fig:immersion-approaches}), although this is a rare achievement and not a required result of design by immersion. Design by immersion typically leads to a drift of (some or all) involved researchers within this knowledge space (as discussed in~\cite{HinrichsEtAl2017}). We do not suggest that there is a correct trajectory to take, this depends on the project and the people involved. However, an awareness of this knowledge space and where one would locate oneself can help reflection on collaborative practices and design activities already taking place. We will identify how we have ``drifted'' through this knowledge space as we discuss our case studies.

The case study descriptions that follow culminate in a series of descriptive tables which can be used as a starting point for new projects. Table 1 lists activities that can benefit a research project. Table 2 provides guidance for understanding the stages of research for those activities. Table 3 describes possible themes and reflective questions to consider, and Section~\ref{sec:challenge-outlook} describes potential problems to watch out for.

\section{Immersion Case Studies}
\label{sec:immersion}
Our case studies cover a wide variety of domains, including computational linguistics, medicine, literary analysis, transport studies and chemistry. While the literature lacks a characterization of design by immersion, we draw on a number of existing studies that exemplify immersion~\cite{IsenbergEtAl2008,BrehmerEtAl2014,WoodEtAl2014,HinrichsEtAl2016,BradleyEtAl2016}. This is not intended as an exhaustive characterization of immersion but rather to illustrate how the ideas in this paper arose from independent places and projects, while capturing the richness and multifacted nature of design by immersion. We use each case study as a way to reveal transdisciplinary activities using \textit{Letter-\#} to refer to the activities in Table~\ref{table:activities}. These activities have been grouped as \textbf{d}ata analysis, \textbf{s}tudy methods, \textbf{p}rototyping, \textbf{l}earning about the other domain, and \textbf{c}ommunicating across domains, though some could be classified in multiple categories. Some activities appeared across multiple case studies as highlighted in Table~\ref{table:activities}, so we focus on detailing a subset associated with each case study.

\subsection{Case Study \#1: Apprenticeship} 
One approach to design by immersion is for a visualization researcher to participate in domain activities (an upward movement in Figure~\ref{fig:immersion-approaches}).

\textbf{Context:} Collins \etal\cite{Collins2009} recount a visualization researcher using immersive observation to become an apprentice in the context of computational linguistics, and thus gain first-hand experience in statistical machine translation. In this case, the visualization researcher's first-hand experiences in the target domain provided a contextualized understanding of it, which informed the visualization design process. The researcher's experiences were a combination of being an immersed intern in the team and consciously exploring the current use of visualization like a type of qualitative pre-design study.

\textbf{Activities and Design Discussion:}
In this case, the lead visualization researcher had some prior knowledge of computational linguistics. The domain expert collaborators worked specifically on statistical machine translation and trained the visualization researcher to carry out common domain analysis tasks (L-4). The visualization researcher subsequently engaged in independent analysis with standard tools, generated his own domain-specific findings (\textit{D-1}, \personal, \Active), and validated his findings with domain experts. By using pre-existing domain knowledge, the immersive researcher was able to start the visualization ideation process early, sharing visualization sketches with computational linguistics experts on a daily basis (\textit{P-3}, \communal). Through shared work environments and meetings, this period involved rapid, collaborative design iterations tightly coupled to informal domain expert feedback (\textit{P-4}), quickly converging on the design of Bubble Sets (\emergent)~\cite{Collins2009}.
During these experiences, the immersed researcher kept a journal, which the research team subsequently leveraged to understand the domain's data and work practices (\textit{S-2}). 

\textbf{Benefits and Impact:}
Insights into the domain problem went beyond those elicited through initial interviews with domain experts. Through experiencing the analysis process first-hand, the visualization researcher was better equipped to identify opportunities where interaction design and visualization could improve the workflow. This work highlights how, through immersion, visualization researchers can engage in independent domain-specific data analysis to gain their own understanding of the target domain and its processes (\textit{D-1}). 

\subsection{Case Study \#2: In-situ Simulation} 
Here we illustrate immersion from both directions, achieved through both visualization researchers and medical experts brainstorming together to derive novel immersive methods.

\textbf{Context:} Some domains, such as medicine, require particular sensitivity and one might think that design by immersion would not be an option. However, the medical domain is one in which domain-specific technology design may be particularly important. 

\textbf{Activities and Design Discussion:} First, the medical collaborator spent considerable time with the visualization group to understand the importance of pre-design empirical work~\cite{BrehmerEtAl2014} (\textit{S-4}, \communal). Next, members of the visualization team observed medical experts in-situ, as in job shadowing, a common learning practice in the medical field (\textit{S-1}). Recognizing that this produced insufficient insight for effective design, the team brainstormed together alternative methods (\textit{C-5}, \communal, \emergent)
 to address the challenge  of enabling in-situ interviews with internists (doctors who consult on internal medical problems), without putting a strain on their already busy and high-pressure workdays. With the medical collaborator and considerable advice from an ethics board (\communal), the visualization researchers designed what was essentially an in-situ interview (\textit{S-3}) where the doctors gave a medical consult in the context of their working environment, minimizing their time  commitment and maximizing the potential of observing their diagnostic process in a close-to-real situation~\cite{BrehmerEtAl2014,ZukPhDThesis}. For the internists, who had agreed to be approached for a consult on pulmonary embolism (PE), the visualization researcher could approach them in the hospital halls, and in a manner similar to how one doctor asks another, ask for a consult on a (non-existent) PE patient (\personal, \Active). The visualization researcher could then, through this simulated consultation, gain a deeper understanding of doctors' diagnostic practices (\textit{L-5}).

\textbf{Benefits and Impact:} Building on extensive transdisciplinary collaboration, the team designed software based on the doctors' own processes (\emergent), using Bayesian reasoning to provide support and revealing data uncertainty when necessary. In a pre-clinical trial comparing the use of this software to a 20~minute refresher lecture, 19~out of 20~doctors did equivalently or better. Now, a professionally programmed version is in clinical trial.

\subsection{Case Study \#3: Immersion in Visualization}
\label{sec:bradley}
Researchers from other domains can immerse themselves in visualization resulting in a rightward movement in the  space of Figure~\ref{fig:immersion-approaches}. 

\textbf{Context:}
This case study, reported by Bradley et al.~\cite{BradleyEtAl2016}, shows the humanities embracing both visualization and the visualization community's methodology. The authors identify a gap in existing visualization techniques for language, introducing a vector space model (L-DNA) to address it. They evaluate L-DNA with a diverse group of participants to highlight the technique's potential and critique its design. 

\textbf{Activities and Design Discussion:} The lead author, Bradley, originally sought out experts in visualization because he had produced mathematical work that his primary domain of English literature was incapable of evaluating. He found value in learning about the visualization process (\textit{L-1}) because it allowed him to query his own research in new and interesting ways, and convinced him to pursue work as a dual citizen in both research domains. 

In this instance, the author was not formally trained in mathematics, and needed a domain expert to help translate mathematical ideas into readable equations. This was a unique situation for both scholars as they learned the language and symbols of each other's domains (\textit{C-4}, \communal, \personal). The openness of the visualization experts to embrace ideas that had their conception in another discipline was a necessary part of this collaboration (\textit{C-1}). One of the difficulties of immersion in a foreign domain is that lack of knowledge of prior work and terminology make it difficult to communicate with domain experts, even if the ideas are novel and relevant. Quite a lot of work was done in translation (\Active), essentially defining common terminology to support conversation. By training, an English scholar's view of language is fundamentally different than those in technical fields, so as conversations became deeper and more theoretical, there was an inverse relationship with the time it took to agree on terminology. While this process was sometimes described as frustrating, it was also exciting in that both parties felt that they were learning from each other while approaching a common goal. This type of immersion can be challenging because it is dependent on both parties being motivated by a process that can be confusing and time consuming. 

\textbf{Benefits and Impact:}
All researchers reported learning new ways of thinking about their own domains (\emergent) by considering how to describe their own tacit knowledge in ways that their colleagues could understand. The willingness of one researcher to immerse themselves fully in the domain of their colleagues led to developments that enabled solving problems in both domains (\emergent) and resulted in the publication of a new method for representing words and documents in a reversible vector space.

\subsection{Case Study \#4: Reciprocal Immersion}
\label{sec:hinrichs}
Design by immersion can also take on a reciprocal character where visualization and domain experts immerse themselves in each others' research approaches and practices, resulting in a vertical upward movement of the visualization researcher in the expertise space and a rightward movement of the domain expert (see Figure~\ref{fig:immersion-approaches}). 

\textbf{Context:} Hinrichs \etal\cite{HinrichsEtAl2016} worked at the intersection of visualization and literary studies to analyze the characteristics of a largely unknown collection of science fiction anthologies. In this project, initial research questions were vague and open-ended due to the underexplored character of the literary collection in focus. The data collection, analysis, and accompanying visualizations evolved over the course of the collaboration (\textit{P-1}) and were shaped by the design by immersion practices that the team established (\emergent).

\textbf{Activities and Design Discussion:} Throughout the project, team members from literary studies and visualization independently engaged with the collection and its evolving data using their own domain specific approaches. The literary scholars engaged in archival work and the classification of the science fiction anthologies from a literary perspective while the visualization expert, even before concrete data was available, developed sketches of ideas (on paper and in computational form) for a visualization system that could facilitate the analysis of the anthologies from different perspectives (\textit{P-1,P-3}). As the project took place across two continents, the visualization researcher could not be involved in the archival work or classification process, or directly observe the practices of the literary scholars first-hand. Immersion in each others' research practices happened instead through frequent online discussions (\textit{C-3}) that helped form joint transdisciplinary perspectives on the project (\communal) that would ultimately facilitate contributions to literary studies~\cite{ForliniEtAl2018}, visualization~\cite{HinrichsEtAl2016} and beyond~\cite{Hinrichs_2018a}. Here, the ever evolving visualization sketches were found to be a central point as they became mediators between the two disciplines in that they exposed certain ideas and assumptions from both a literary and visualization perspective. For example, literary scholars rejected certain visualization ideas that would shape the interpretation of the collected data in unwanted ways (\textit{P-2}). The visualization sketches raised questions that informed new angles to the ongoing data collection and archival work (\textit{D-2}). More specifically, documenting these discussions alongside all visualization sketches helped the researchers reflect on their collaborative process (\textit{S-2}). While immersion initially took place in the form of team meetings, the researchers also started to immerse themselves in the ``other'' domain, by reading and discussing relevant literature (\textit{L-1}) and by co-authoring articles and participating in each others' conferences (\personal, \Active). Throughout the project, the team members supported immersion in each other's fields. More generally, domain experts can help visualization researchers immerse themselves and visualization researchers can reciprocate.

\textbf{Benefits and Impact:} The visualization and literary researchers evolved concurrently and influenced one another significantly. This transdisciplinary collaboration enabled by design by immersion has fundamentally shaped the research team and enriched but also changed each team member's perspective on their own field. They have ``drifted'' away from their own domain toward those of their collaborative partners~\cite{HinrichsEtAl2017}. This case study shows that design by immersion as a process is well suited to work in the context of evolving collaborative visualization-driven research projects involving disciplines where research goals and questions are still in-flux (\textit{P-1}).

\subsection{Case Study \#5: Dual Citizens}
The deepest form of immersion occurs where a researcher is \textbf{both} a visualization expert and a domain expert --- a dual citizen occupying the top-right space in Figure 1. In this case study, reported by Wood et al.~\cite{WoodEtAl2014}, we consider problem solving by researchers with expertise in both visualization design and transport studies.

\textbf{Context:} 
The problem motivating this collaboration between academics and transport planners was the efficient provision and expansion of a major bicycle-share scheme in London, UK. Those directly involved included transport authority analytics experts, transport policy managers, and `dual citizen' academics with both visualization and geography/transport backgrounds (\textit{P-1}).

\textbf{Activities and Design Discussion:}
A key activity in the initial discussion between transport authority members and academics was the establishment of \textit{trust} (\personal)---a genuine belief among participants that the investment involved in collaboration will be beneficial in addressing the motivating problem. Having dual citizens involved helped by demonstrating a commitment to the domain  (\textit{C-1}, \personal, \Active), providing a common (transport-related) framework for dialogue (\communal, \textit{C-2}), and demonstrating recognizable expertise. This accelerated the transition to the analytics phase of the collaboration (\Active) where data and visualization provided the mediation artifacts for discussion (\textit{P-4}). In parallel with this collaborative activity between academics and transport authority members, dual citizenship supported participation in the academic transport studies community including methodological innovation in the use of transport visual analytics (\emergent). Importantly, members of the transport authority participated in joint publication and presentation in the academic transport studies sector (\textit{S-4}, \personal, \communal, \Active, \emergent) --- something unlikely to have occurred without the dual citizenship of some of the participants.

\textbf{Benefits and Impact:}
Immersion via dual citizenship can speed up the process of collaboration, especially at the trust establishment phase of a project. It also offered a wider range of opportunities for impact (operational, policy, academic visualization, and transport studies) than might otherwise have been the case. It provided a structure for the PhD of one of the participants and career development for another (\personal). It assisted in the use of visualization as mediation between the analytics and senior policy members of the transport authority. This deeper immersion also opened up other parallel channels of mutually supporting activity in the arts and museum sectors~\cite{WoodEtAl2014}.

\subsection{Case Study \#6: Dual Immersion} 
\label{sec:RAMT}
Design by immersion was used as part of a collaboration to create novel visualizations in the chemistry domain ~\cite{Codorniu_2014,Codorniu_2015}. Here, we discuss how dual immersion was used to achieve an effective visualization design.

\textbf{Context:}
The team comprised a visualization researcher, two chemical researchers, and a researcher immersed in both visualization and chemistry. At the start of the collaboration, the immersive researcher was developing both his visualization and chemistry expertise in order to become a dual citizen of both research communities. We focus on the impact his immersion in chemistry had on visualization design. 

The immersive researcher initially saw an opportunity to address chemistry data analysis challenges through visualization, and the collaborative team evolved because he actively chose to enter into these complicated research spaces. The team was originally what Kirby and Meyer~\cite{KirbyMeyer-2013} call a \textit{multidisciplinary team}; trying to solve a research challenge in one domain by leveraging multiple domains such that team members had clear roles. However, the distinction between visualization researcher and domain expert became fluid. By the end of the collaboration, all team members warranted recognition for having helped advance both visualization and chemistry. It is important to note that transdisciplinary activity need not apply to full teams to be useful. Here, one researcher on the team acted in a transdisciplinary context, and it helped shape the way the entire team worked together.

\textbf{Activities and Design Discussion:}
 Having an immersive researcher in both chemistry and visualization enabled knowledge development within the team. Leveraging his existing chemistry knowledge, the immersive researcher familiarized himself with project-relevant chemistry literature (\textit{L-1}). Coordination and cross-disciplinary learning were achieved through informal peer-to-peer discussions (\textit{C-3}) facilitated by shared domain knowledge (\textit{C-2}) and through casual learning opportunities facilitated by sharing office space and informal observations (\textit{S-1}, \communal). He also engaged in more formal one-on-one and collaborative discussions with the domain researchers about the chemical processes they were studying, chemical research challenges, and potential visualizations. In particular, one of the domain researchers invited him to collaboratively analyze some chemistry data (\textit{D-3}, \communal, \personal, \Active), which strengthened their collaborative relationship and enriched the immersive researcher's understanding of the data analysis challenges. On the visualization side, the immersive researcher regularly discussed visualizations and the corresponding chemistry with the visualization researcher, acting as a translator between the visualization and chemistry experts (\textit{C-4}). During the collaboration, the immersive researcher kept sketches, personal notes and prototypes which were valuable in reflecting on the design process (\textit{S-2}). Furthermore, he started to self-critique his designs from \textit{both} the chemistry and visualization perspectives as the project and his immersion progressed (\textit{P-5}). The collaboration resulted in several visualizations, for example Radially-Angularly Mapped Trajectory (RAMT) plots ~\cite{Codorniu_2014, Codorniu_2015}. While RAMT plots resolve a specific domain analysis challenge, they incorporate additional domain concepts that were not directly connected to the domain challenge nor the chemistry researchers' data characterizations. Instead, these domain concepts emerged as relevant during the design process. Immersion helped the researchers gain a broader exposure to domain concepts (\textit{L-2}), which they leveraged to: develop domain-inspired visual encodings that explicitly incorporated domain knowledge (\textit{P-2}, \emergent), and reveal domain-based design considerations for visualizations (\textit{L-3}, \emergent).

\textbf{Benefits and Impact:}
The project resulted in novel visualizations that advanced chemical understanding, and yielded publishable insights~\cite{Codorniu_2014,Codorniu_2015}. It revealed design considerations for future molecular visualization. The resulting chemistry visualizations also raised visualization questions, and inspired subsequent work exploring emphasis in information visualization~\cite{HallEtAl2016}.

%% file: 04_opportunities.tex
\section{Relating Immersion to Design Opportunities}
\label{sec:opporunities}
To support other researchers engaging in design by immersion, we reflected on our shared experiences using six case studies from our recent problem-driven visualization work. In all of the case studies, researchers immersed themselves to varying extents in multiple domains through different strategies. Collectively, our case studies point out the benefits of this process. We have identified three opportunities. (1)~Design by immersion enriches domain understanding through personal domain experiences. (2) It facilitates the exploration of new domain-inspired research and design spaces, and (3) it promotes the building of mutual and productive transdisciplinary relationships that advance all disciplines involved.
Table~\ref{table:activities} connects these opportunities to the design activities detailed in the previous section.

\subsection{Enrich Domain Understanding through Personal Domain Experiences}
\label{sec:domain-des}
Gaining an understanding of a target domain plays a central role in models of problem-driven visualization design (\eg~\cite{Munzner2009, SedlmairEtAl2012, McKennaEtAl2014}).  Immersion enables researchers to leverage a variety of activities to enrich their understanding of a target domain. What makes design by immersion distinct is the personal component of how a researcher understands a target domain.
Researchers can engage in collaborative data analysis with domain experts (\textit{D-3}), or analyze domain data independently to gain first-hand experiences with existing tools (\textit{D-1}). An immersed researcher can unobtrusively observe domain experts in situ (\textit{S-1}), or directly engage in peer-to-peer communication with them about both the domain science and visualizations (\textit{C-3}).  For additional perspectives on the domain, an immersed researcher can consult domain literature (\textit{L-1}). Similarly, previous work has suggested that visualization researchers read domain literature (\eg\cite{Wijk2006,SedlmairEtAl2012}), and immersion will help them gain more from this activity. Through immersion, researchers also gain deeper exposure to the concepts and problems of other domains beyond visualization challenges (\textit{L-2}). In each of these activities, the researcher either engages directly with domain material, or explores the domain through personal exchanges with domain experts that emphasize peer-to-peer relationships. By participating in the target domain, the researcher confronts domain challenges and creates actionable mental models of the domain. As a researcher starts to design visualizations following (or as part of) immersion, the goal is to develop a richer personal understanding of the target domain.

Researchers can also use immersion-supported activities to evaluate domain descriptions (\eg \textit{S-1, D-3, C-3}), particularly when they leverage their domain experiences and connections to access new external domain experts for validation purposes. In the context of Munzner's nested model~\cite{Munzner2009}, immersion can facilitate both exploratory and summative evaluations---pre-design studies aimed at gaining a understanding of a domain situation and post-design evaluations aimed at assessing research output quality in the context of the domain situation.
McKenna \etal\cite{McKennaEtAl2014} also distinguish methods that generate an understanding of a domain task and those that evaluate an existing characterization of a task. Immersion empowers researchers to enrich their domain understanding through personal domain experiences, and provides mechanisms for evaluating domain descriptions. Immersion can also potentially support approaches aimed at achieving ``immersive'' tool evaluations (e.g., see ~\cite{ShneidermanBELIV2006}).

\subsection{Explore New Domain Inspired Spaces}
Design by immersion provides an alternative perspective on how researchers can explore visualization solution spaces. For example, through immersion, researchers can realize visual encodings that explicitly leverage domain knowledge and practices beyond those captured by domain problem characterization and subsequent abstraction (\textit{P-2}), in part due to their broader exposure to domain concepts (\textit{L-2}). While previous work has suggested that deriving new data types is part of visualization design\cite{McKennaEtAl2014}, an immersed researcher, with both domain experience and visualization knowledge, is particularly well positioned to manipulate and extend datasets in meaningful ways (\textit{D-2}).
Wood et al. detail an example of this in their work~\cite{WoodEtAl2014}. In turn, immersion can help meet the design guidelines of Wong et al.~\cite{WongBELIV2018} that outline how researchers designing visualizations for domain experts should use domain terminology in their designs.

Immersion also opens up alternative design paths. Immersed researchers can concurrently characterize new domain problems and ideate by leveraging their existing experiences and knowledge of the domain (\textit{P-3}). In contrast, previous work (\eg~\cite{Munzner2009,SedlmairEtAl2012,McKennaEtAl2014}) takes the perspective that visualization projects involve an initially ordered progression from domain characterization to ideation. Design by immersion involves visualization researchers iterating frequently and collaboratively on designs with domain experts (\textit{P-4}) so that visualizations are intertwined combinations of input from domain expert collaborators, visualization and domain influences, and prototype development, aligning well with the build-intervene-evaluate cycles of ADR~\cite{SeinEtAl2011}. The tight coupling of visualization and domain during design by immersion also empowers researchers to design visualizations in the context of evolving collaborative projects (\textit{P-1}), and potentially when domain needs are open-ended~\cite{HinrichsEtAl2016}. In contrast, previous work~\cite{Munzner2009,SedlmairEtAl2012,McKennaEtAl2014} generally emphasizes visualization design scenarios where domain experts have relatively stable tasks and workflows. 

Immersion also changes how visualization researchers consider and explore solution spaces as they generate visualization designs. Immersion enables the researcher to critique their designs from \textbf{both} the domain and visualization perspectives (\textit{P-5}). This dual self-critiquing is complementary to external critiques from domain experts. Self-critiquing from the domain perspective is similar to Neustaedter and Sengers' autobiographical design process~\cite{Neustaedter2012}.  Autobiographical design is a type of design research in HCI that relies on ``extensive, genuine usage by those creating or building the system''~\cite[p.514]{Neustaedter2012}. They advocate that researchers leveraging autobiographical design should keep formal records to facilitate reflection on their design processes. We recommend that immersed researchers should similarly keep their own records if they want to reflect on their design processes following an immersive project, particularly given that design by immersion generally involves blurred roles. Documenting the process both helps reflecting on the domain and navigating the design process ~\cite{HinrichsEtAl2016} (\textit{S-2}).

Design by immersion can potentially reveal new problem spaces through mutual shaping. Mutual shaping is a principle of ADR~\cite{SeinEtAl2011,McCurdyEtAl2016}, and refers to how members of the design team come from differing backgrounds, but learn from each other and shape one another's ideas. In particular, ``Through close collaboration the team members learn about each other's expertise, sometimes offering valuable insight into another member's primary research domain''~\cite[p.514]{McCurdyEtAl2016}. We experienced this during two of the case studies (RAMT plots and the L-DNA project~\cite{BradleyEtAl2016}). Outsiders are well positioned to question assumptions and reveal new challenges that the other domain has not considered.

By having both visualization and domain knowledge, an immersive designer brings broader knowledge, perspectives, and skills when addressing a domain's visualization challenges and critiquing visualizations. The close connections established between the researcher and their domain expert collaborators pushes the design process away from a cyclic interaction. Knowing the target domain and exploring the visualization space become intertwined. 

\subsection{Build Interdisciplinary Relationships}
Through immersion, visualization researchers share more in common with their domain expert counterparts, facilitating greater rapport among collaborators. By embracing design by immersion, a visualization researcher can utilize language that resonates with their domain collaborators (\textit{C-1}); the value of which has been noted by others\cite{ShneidermanBELIV2006,SimonEurovis2015}. Moreover, akin to Liaisons~\cite{SimonEurovis2015}, immersed researchers can relate to domain experts through common knowledge or experiences (\textit{C-2}), engage in informal peer-to-peer communication (\textit{C-3}),  collaboratively analyze data with domain experts (\textit{D-3}), and translate ideas for non-immersed members of a design team (\textit{C-4}). These activities, in addition to helping establish domain understanding, help immersive designers to create cohesive teams. Greater rapport and cohesion can establish the trust and openness necessary for effective collaborations.  Furthermore, strong relationships with domain experts offer new opportunities, such as access to an extended network of domain experts, and provide a firm basis for engaging in more ambitious future work. While Sedlmair \etal\cite{SedlmairEtAl2012} identify poor rapport with collaborators as a potential pitfall for design studies, the visualization literature generally does not emphasize that building bridges between visualization and other domains requires visualization researchers to build effective personal relationships with researchers in other domains. Design by immersion can help visualization researchers build these personal relationships.

%% file: 05_immersion-process.tex
\section{How to Immerse Yourself}
\label{sec:process}
Immersing oneself and engaging in design by immersion are two interconnected research processes. The activities in Table~\ref{table:activities} are intended as practical approaches to facilitate the construction of customized immersion trajectories. 
There are many activities in Table~\ref{table:activities} and not every design by immersion project will use all possible activities. In fact, one of the challenges of design by immersion is that there are many possible immersion trajectories, and one needs to navigate these possibilities while remaining sensitive to one's immersion domain. What is appropriate in the context of one domain may not be suitable for another domain. There is also the challenge of deciding when to stop; immersion trajectories do not necessarily need to culminate in dual citizenship. In light of this complexity, we aim to empower people to develop their own trajectories.
To this end, we first illustrate the diversity of possible trajectories leveraging our case studies, and then provide a framework for constructing immersion trajectories. 

\subsection{Example Trajectories}
\label{sec:example-traj}
Immersion trajectories may vary considerably depending on the project, team members, and domains. To illustrate the diversity of possible trajectories, we review the trajectories for \textbf{Case Studies \#4} and \textbf{\#6}.

In \textbf{Case Study \#4}, the two immersed researchers originated from literary studies and visualization. They facilitated each other's immersion, which was shaped by the idea of developing visualizations in an evolving transdisciplinary project (\textit{P-1}). The immersion trajectories involved data-driven activities (domain-specific analysis \textit{D-1}, deriving new data \textit{D-2}, and collaborative data analysis using methods from literary studies and visualization \textit{D-3}). Visualization and prototyping followed a speculative approach, that paralleled these data-driven activities (\textit{P-3}), in particular: developing visual encodings that explicitly took inspiration from domain knowledge and practices (\textit{P-2}), and iterating rapidly and collaboratively on designs by leveraging informal domain expert feedback (\textit{P-4}). Key to the immersion for both researchers were frequent discussions of emerging insights from archival work and visualization (\textit{C-3}). Current visualization prototypes (on paper and digital) as manifestations of insights, assumptions and questions became the centre of these discussions that also included design critiques from visualization and literary studies perspectives  (\textit{P-5}). While immersion happened as part of these discussions, documenting them enabled later reflections on how immersion had affected the course of the project and project outcomes as well as the researchers themselves (\textit{S-2}).

In \textbf{Case Study \#6}, the immersed chemistry researcher's trajectory involved data-driven activities, learning about the visualization domain, communicating with collaborators, and study methods, some of which are domain expert analogs of activities in Table~\ref{table:activities}. For example, the researcher took a course in visualization, read introductory visualization literature (\textit{L-1}), and then undertook independent design sketching and digital prototyping (\textit{D-1}). He started participating in a visualization research group, which involved attending the group's meetings and engaging in informal peer-to-peer discussions about visualization (\textit{C-3}). Through this engagement, he was able to observe visualization design practices unobtrusively in-situ (\textit{S-1}), in addition to critiquing and designing visualizations collaboratively with visualization researchers (\textit{D-3}). Longer term, he attended visualization conferences, and collaborated with visualization researchers on visualization projects beyond chemistry-specific applications.

The differences in these trajectories highlight how specific low-level characteristics of immersion can differ between projects. Therefore, we have generated a framework with several sets of questions to help guide researchers as they construct their own immersion trajectories.

\subsection{Constructing an Immersion Trajectory}
\label{sec:construct-traj}
Researchers need to carefully consider and construct their own immersion trajectories while respecting the specifics of their own interests and personalities, immersion domain and research goals, as well as collaborators. Design by immersion is not a cure all. However, in our experience, it tends to lead to rich insights and changes in both personal and collaborative perspectives, advancing all disciplines involved. We provide a framework for constructing immersion trajectories in order to enable other researchers (within and outside of the visualization domain) to leverage design by immersion. 

\subsubsection{Considering Your Own Interests \& Contextual Factors}
Researchers choosing to start design by immersion projects should carefully consider why and how they will build transdisciplinary relationships, and, thus, surround themselves with a support group. 

\textbf{What are your own interests?} --- It can be important to reflect on your own interests and what draws you toward this type of approach. Maybe your profile is already interdisciplinary and you wish to experience work in another domain in more detail. Maybe you wish to incorporate another domain's perspective as you feel it may deliver key aspects to your project and/or research interests. In any case, it can be valuable to consider personal goals. What would you like to get out of the transdisciplinary collaboration? What do you think you might contribute to the collaboration and/or immersion?

\begin{table}[b!]
\vspace{-0.8em}
\caption{An example immersion trajectory.}
\vspace{-1.8em}
\begin{center}
\setlength\extrarowheight{3pt}
\begin{tabular}[t]{m{0.24\columnwidth}@{\hskip0pt}m{0.70\columnwidth}@{\hskip0pt}}
\toprule
\centering \textbf{Stages}& \centering \textbf{Immersion Trajectory Activities} \tabularnewline \midrule
\centering \textbf{Stage \#1:} Establishing basic \\domain \\knowledge &
\begin{tabular}{ m{0.65\columnwidth}}
Take introductory course(s) in a domain if you have no prior education or experience with that domain
\tabularnewline \arrayrulecolor{lightgray} \midrule
Find a domain expert who is willing to support the immersion process
\tabularnewline \midrule
Read introductory background domain \newline literature
\tabularnewline \midrule
Start attending group meetings of domain \newline researchers
\end{tabular}
\tabularnewline \midrule
\centering  \textbf{Stage \#2:} \\ Gaining \\domain\\ research exposure &
\begin{tabular}[t]{ m{0.65\columnwidth}}
Read domain research literature
\tabularnewline \arrayrulecolor{lightgray} \midrule
Start working co-located with domain experts
\tabularnewline \midrule
Undertake independent domain-specific data analysis
\tabularnewline \midrule
Work collaboratively with domain experts to tackle domain questions and data analysis \newline challenges
\end{tabular}
\tabularnewline \midrule
\centering \textbf{Stage \#3:} \\ Establishing \\broader\\ domain reputation &
\begin{tabular}[t]{ m{0.65\columnwidth}}
Collaboratively engage with domain experts to prepare a domain-specific paper related to your work with them
\tabularnewline \arrayrulecolor{lightgray} \midrule
Attend domain-specific events \newline (\eg a conference) 
\end{tabular}
\tabularnewline \bottomrule
\end{tabular}
\end{center}
\label{table:immersion-process} 
\vspace{-1.2em}
\end{table}

\textbf{Who will be your critical domain supporter(s)?} --- The immersion process will be easier if you have at least one close domain supporter who is willing to invest themselves in the immersion process and make long-term commitments to your work. This strong supporter will likely be the domain partner for your immersive project, so that there is an incentive for them to make long-term commitments. One key feature of this supporter should be an openness in facilitating your immersion, potentially by allowing you access to their work processes and even their work environment. Much of the guidance from Sedlmair \etal\cite{SedlmairEtAl2012} about selecting potential collaborators applies to finding an appropriate domain supporter. Ideally, as your immersion progresses, you should cultivate your connections to the domain community, and develop a succession plan, in order both to raise up supporters for future projects, and to develop contingency connections in the event that your current supporter can no longer be a part of the project as a result of, for example, relocation or institutional realignment. 

\textbf{Who will be your supporting cast?} --- Also consider the broader types of support. Are you cultivating casual relationships with other domain experts? These relationships do not need to be as strong or time intensive as the relationship with your critical domain supporter, but having a broader network will make it easier to get additional perspectives on the domain, related concepts, and challenges.

\begin{table*}[h!]
\caption{Questions for immersive experiences as inspired by experiential learning literature. Concrete experience themes are quoted directly from Morris' recent review on concrete experience and experiential learning \cite[p.4]{MorrisILE2019}.}
\vspace{-0.8em}
\begin{center}
\setlength\extrarowheight{2pt}
\begin{tabular} { m{0.35\textwidth}m{0.55\textwidth} }
\toprule
\centering \textbf{Concrete Experience Themes\cite{MorrisILE2019}} & \centering \textbf{Immersive Experience Questions} \tabularnewline \midrule
\textit{Hands on participation: Involved, active, participants in real-world uncontrived scenarios} &
How will you be active and participating in the domain? \newline
If the real-world activity was modified to make it accessible to you (the visualization researcher), how specifically has the activity changed? How might those changes affect your understandings of the domain?
\tabularnewline \midrule
\textit{Situated in context: Rich contextual information} &
What are the contextual factors influencing the activity and your participation (for example, social factors)? \newline
Which contextual factors does your design respect, and which are you disrupting?
\tabularnewline \midrule
\textit{Critical reflection: Contextual conditions of problem considered} &
How will you reflect on your activities and experiences to develop your own mental models and domain understanding?
\tabularnewline \midrule
\textit{Purposeful~\textemdash~pragmatic: Solving real-world problems} &
How could the activity (as completed by you or a domain expert) help advance the domain? 
\tabularnewline \midrule
\textit{Risk~\textemdash~novel problems: Out of comfort zone: temporary destabilization}&
What are the specific risks associated with you activity? \newline 
What is your own risk tolerance?
\tabularnewline \bottomrule
\end{tabular}
\end{center}
\label{table:experiential} 
\vspace{-2.5em}
\end{table*}

\subsubsection{Selecting Activities}

We suggest breaking down immersion trajectories into three stages that may be sequentially arranged or re-visited iteratively, depending on the project. These stages are: 
\begin{description}
\vspace{-0.3em}
\item[Stage \#1] Establishing basic domain knowledge
\vspace{-0.6em}
\item[Stage \#2] Gaining domain research exposure
\vspace{-0.6em}
\item[Stage \#3] Establishing a broader domain reputation
\vspace{-0.3em}
\end{description}
These stages are outlined in the brief immersion trajectory exemplar in Table~\ref{table:immersion-process}. The three stages are transferable across immersion trajectories, although the extent of Stage \#3 is quite variable. Stage \#3 represents a significant time investment, so some visualization researchers may choose to stop at Stage \#2. However, establishing a broader domain reputation has potential benefits. For example, it might be helpful for recruiting domain experts as participants for studies. 

Researchers can populate the three stages using activities from Table~\ref{table:activities}, as illustrated in Table~\ref{table:immersion-process}. We have ordered the activities in Table~\ref{table:immersion-process} according to one possible immersion trajectory. They are not numbered because we envision that some of them can occur concurrently, and the activities may be reordered depending on a particular researcher's preferences or target domain. Some steps could occur as part of multiple stages. For example, \textit{starting to attend group meetings} could be part of Stage \#1 or Stage \#2. Table~\ref{table:immersion-process} is meant as a starting template, and can be adapted according to the activities chosen from Table~\ref{table:activities}.

When selecting activities from Table~\ref{table:activities}, we recommend researchers consider the themes and questions provided in Table~\ref{table:experiential}. In the introduction to this paper, we highlighted that design by immersion relates to experiential learning, and we build on this idea here to inform the planning of immersion trajectories. More specifically, Morris reviewed the education literature to understand the nature of concrete experience, which is foundational to experiential learning, and extracted five related themes~\cite{MorrisILE2019} as provided in Table~\ref{table:experiential}. We have used these themes as a starting point for constructing the questions in Table~\ref{table:experiential} to help guide researchers as they engage in immersion and select corresponding activities. Our questions in Table~\ref{table:experiential} are provisional, and we expect that they will be revised and expanded by future work. In general, educational literature on experiential learning could help inform and enhance the community's understanding of what constitutes good domain experiences for achieving design by immersion.

%% file: 07_outlook.tex
\section{Design by Immersion Challenges}
\label{sec:challenge-outlook}
While design by immersion provides opportunities, it also has its challenges, which we have gathered from our case studies and experiences.

\textbf{Time Commitment} --- Immersion can place substantial demands on a researcher's time. 
This is not a unique characteristic of this approach. As Munzner~\cite[p.927]{Munzner2009} states, ``The domain problem characterization stage is both difficult and time consuming to do properly.'' What is particularly demanding is the double counting of certain activities, such as engaging and keeping up-to-date with two distinct bodies of literature. However, we highlight again, that design by immersion allows for different levels of trajectories in the immersive skill space (see Figure~\ref{fig:immersion-approaches}), and becoming a ``dual citizen'' is not necessarily the goal. 
This being said, design by immersion does align with recent trends in the visualization community for researchers to specialize in particular application areas, such as BioVis, SecurityVis, or Digital Humanities. 

\textbf{Keeping Track of the Process \& the Process On-Track} --- Design by immersion is suited to visualization design in the context of evolving, open-ended domain projects. This is an exciting opportunity, but also a potential challenge as keeping a focus can be difficult as new perspectives and new questions emerge. As a project evolves, many ideas will crop up, some to be immediately explored while others may be interesting but ``better left for another day.''

\textbf{Navigating Research Identities} --- Researchers may also find it challenging to navigate multiple research identities, specifically their visualization identities and their domain identities. Sticking too tightly with one's ``home'' domain identity can make immersion and insights less enriching. However, maintaining separate identities is important. Failing to do so could mean that a researcher loses the ability to interact appropriately with one of their communities, 
and that the community no longer treats the researcher as an equal citizen.
An awareness of the tension of multiple identities can help one find a suitable path depending on one's own interests and goals.

\textbf{Translating Ideas from one Domain to the Other} --- Translation of ideas from one domain to the other remains challenging. 
It is a skill that derives from experience and from embracing the epistemological concerns of the domains in which the researcher is immersed. Moreover, we note that it requires a certain degree of humility, of being willing to acknowledge the limits of one's own understanding. 

\textbf{Systemic Barriers} --- Researchers leveraging design by immersion may encounter systemic barriers related to its transdisciplinary nature. For example, most PhD programs are structured around a single home department. Similarly, funding initiatives often target
particular domain groups (\eg the sciences, social sciences, the humanities, etc.), presenting challenges for transdisciplinary collaborations across non-classical cognate disciplines (e.g., computer science and literary studies). However, it is worth noting that many companies are resource lean, so transdisciplinary work experience and dual citizenship may be desirable. Systemic barriers, like immersion, are sensitive to one's context.

%% file: 08_conclusions.tex
\vspace{-0.4em}
\section{Conclusion}
\label{sec:conclusions}
The visualization community is grappling with increasing multidisciplinarity and the breakdown of the traditional dichotomy between visualization researchers and domain experts. In this paper, we introduced and discussed design by immersion, a progression of themes already present in the visualization literature. Design by immersion is an alternative design approach for problem-driven visualization research, and expands the suite of tools available to the community. We highlighted how immersion supports design activities and provides researchers with opportunities to: (1) enrich domain understanding through personal domain experiences, (2) explore new domain-inspired spaces, and (3) build interdisciplinary relationships. To empower other researchers to take advantage of immersion, we pointed to alternative strategies for achieving immersive goals, related the process to existing design approaches in the literature, and revealed some challenges. We also provided a high-level road map of how visualization researchers can immerse themselves in target domains. The visualization community has been calling for increased application research for some time, and a critical part of achieving this is breaking down the walls between visualization and other domains. Design by immersion helps address these challenges by empowering researchers to explore new transdisciplinary horizons for problem-driven, applied visualization.